\documentclass{jpp}
\usepackage{graphicx}

\usepackage[utf8]{inputenc}
\usepackage[T1]{fontenc}
\usepackage{amsmath}

\usepackage{amssymb}
\usepackage{xcolor}
\usepackage{hyperref}
\hypersetup{
    colorlinks,
    linkcolor=blue,
    citecolor = blue,
    filecolor=red,      
    urlcolor=blue,
    }
\usepackage{caption}

\usepackage[normalem]{ulem}
\usepackage{float}

\title{Firewall effect on electron acceleration by R-waves and parallel electric fields}

\author{Hye Lin Kang\aff{1},
        Young Dae Yoon\aff{1,2}
        \footnote{Present address: Department of Physics and Astronomy, University of California, Los Angeles, CA 90095, USA}
        \corresp{\email{ydyoon@physics.ucla.edu}},
        \\
        Myung-Hoon Cho\aff{3},
   \and Gunsu S. Yun\aff{1,4}
        \corresp{gunsu@postech.ac.kr}}

\affiliation{
\aff{1}Department of Physics, Pohang University of Science and Technology, Republic of Korea
\aff{2}Asia Pacific Center for Theoretical Physics, Pohang, Republic of Korea
\aff{3}Pohang Accelerator Laboratory, Pohang University of Science and Technology, Pohang, Republic of Korea
\aff{4}Division of Advanced Nuclear Engineering, Pohang University of Science and Technology, Pohang, Korea
            }

\begin{document}

\maketitle

\begin{abstract}
We report an unanticipated electron dynamics in a classical setting of a uniform magnetic field, a parallel electric field, and a right-handed circularly polarized wave (R-wave). The setting admits a natural trajectory that a particle accelerated by the electric field reaches a Doppler–shifted cyclotron resonance and becomes trapped in the resonance space. Remarkably, once it becomes resonantly trapped, the electron undergoes reversal of parallel acceleration together with perpendicular energization, despite the parallel electric field remaining constant. This counterintuitive behavior has important implications for particle scattering in various laboratory and space plasmas. Applied to fusion devices, particle-in-cell simulations show that an externally injected R-wave can act as a firewall suppressing further runaway-electron acceleration.
\end{abstract}


\section{Introduction}
Resonant interactions between charged particles and electromagnetic waves govern energetic-particle dynamics in various magnetized plasma environments. In fusion plasmas, controlling  runaway electrons is a central challenge \citep{Breizman2019,Salewski2025}, and wave-induced momentum redirection offers a promising mitigation route \citep{Zhou2011,Guo2018,Yoon2021,Decker2024,Choudhury2025}. Similar resonant mechanisms shape energetic-electron behavior in space, producing solar-wind and flare–accelerated populations \citep{Reames1994,Vocks2005,Lacombe2014,Stansby2016}, auroral emissions \citep{Horne2003,Miyoshi2010,Kasahara2018}, and radiation-belt losses \citep{Inan2003,Sauvaud2008,Compernolle2014,Yoon2020}. Energetic ions in tokamaks \citep{Heidbrink2020,Kirov2024} and stellarators \citep{Siena2020,Paul2022,Paul2023} likewise interact with Alfvénic and ion-cyclotron waves, leading to redistribution and confinement degradation. During magnetic reconnection, waves and energetic particles are simultaneously generated and thus mutually interact \citep{Yoo2024,Park2025}. Many of these interactions occur through Doppler-shifted cyclotron resonance, which drives particle scattering in both quasilinear \citep{Lyons1974,Summers2005,Pokol2008,Liu2018,Guo2018} and single-particle regimes \citep{Bellan2013,Yoon2020,Yoon2021}. 

These interactions in many cases involve an electric field parallel to the background magnetic field. For instance, parallel electric fields are responsible for runaway electron generation in tokamaks \citep{Connor1975,Helander2002,Marshall2019,Lee2024} and are also important during guide-field magnetic reconnection \citep{Schindler1988, Louis2025}. They are also readily observed in Earth's magneto/ionospheres \citep{Gurnett1972, Yeh1981, Falthammar1989, Ergun2002} and in magnetic mirror devices \citep{Geller1974, Burdakov2010}, where plentiful amounts of wave-particle interactions occur.

Previous studies have shown that the coexistence of parallel forces and resonant wave interactions can produce unconventional particle dynamics, mainly in the context of space plasmas. For example, \cite{Kuramitsu2005-PRL, Kuramitsu2005-A&A, Kuramitsu2005-JGR} demonstrated that when an electrostatic field profile maintains cyclotron resonance between a charged particle and a circularly polarized wave, substantial perpendicular energization can occur, a process referred to as "gyroresonant surfing". Similarly, \cite{Omura2007} showed that the combined action of the magnetic mirror force and the whistler mode waves, one class of the R-waves, can lead to efficient electron acceleration through the mechanism known as "relativistic turning acceleration".

Despite these important advances, several limitations remain in the previous studies. \cite{Kuramitsu2005-PRL, Kuramitsu2005-A&A, Kuramitsu2005-JGR} solved the non-relativistic equations of motion and modeled the evolution of the perpendicular velocity, assuming that the parallel velocity is continuously maintained at the Doppler resonance condition by a prescribed electrostatic field profile. \cite{Omura2007} considered only mirror-trapped electrons with sufficiently large perpendicular velocity and whose parallel velocity reach zero. A generalized theory describing the coherent wave-particle interaction under parallel forces is thus warranted.

\begin{figure}
\centering
\includegraphics[width=0.75\columnwidth]{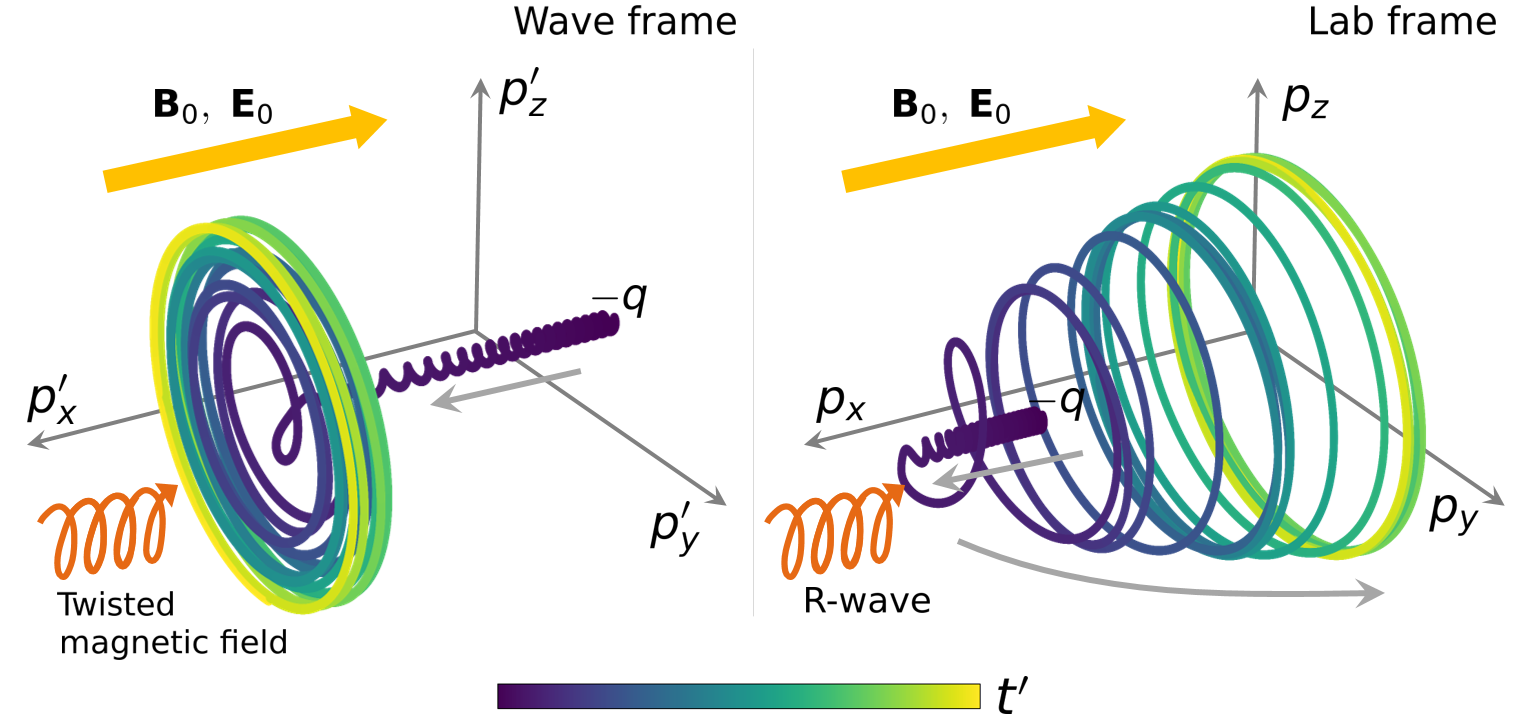}
\caption{\label{fig:schematics} Schematic diagram of an electron trajectory in (left) the wave-frame and (right) the lab-frame momentum space under a uniform magnetic field, parallel electric field, and a right-handed circularly-polarized wave (R-wave).}
\end{figure}

In this paper, we perform a rigorous, exact theoretical treatment of the  fully relativistic interaction between an electron and a right handed circularly polarized wave (R-wave) in a constant parallel electric field by means of a pseudo-potential analysis \citep{Bellan2013}. Far from Doppler-shifted cyclotron resonance, the parallel electric field accelerates the electron towards resonance. For a sufficiently strong wave amplitude the particle may be trapped near resonance, at which point it undergoes acceleration in the direction opposite to its prior motion as well as in the perpendicular direction. This reflection, which is absent in the non-relativistic limit, is schematically shown in Figure \ref{fig:schematics}. Although we focus on interactions between electrons and R-waves, the mechanism is fully generalizable to positively charged particles such as ions and L-waves. This counterintuitive analytical prediction is verified by single-particle simulations and further confirmed by self-consistent particle-in-cell simulations using typical tokamak plasma parameters. We demonstrate that an externally applied wave can act as an acceleration firewall, preventing runaway electrons from accelerating beyond a prescribed momentum, as well as accelerating them in the perpendicular direction. Implications for other systems, including space, astrophysical, and stellarator plasmas, are also discussed. 

\section{Electron motion in a coherent R-wave ($E_{0}=0$)}

To sequentially introduce our analytical derivation, we first consider a negatively charged particle with rest mass $m$, charge $q<0$, and momentum $\mathbf{p}=\gamma m\mathbf{v}$ subject to a coherent right-handed circularly polarized wave (R-wave) and a steady-state background magnetic field directed along the $x$-axis:

\begin{eqnarray}
\mathbf{B} = & & B_{0}\left[\hat{x}+b\left(\hat{y}\sin\left(kx-\omega t\right)+\hat{z}\cos\left(kx-\omega t\right)\right)\right],\nonumber \\
\mathbf{E} = & & -\frac{\omega}{k}\hat{x}\times\mathbf{B},\label{eq:EM fields in lab frame}
\end{eqnarray}
where $b$ is the relative amplitude of the wave magnetic field, which is not necessarily small. The wave angular frequency $\omega$, wavenumber $k$, and $b$ are assumed to be constant. Note that although we consider here a negatively charged particle under an R-wave, the dynamics is identical for a positively charged particle under an L-wave \citep{Yoon2020}. 

In the frame moving with the wave phase velocity $v_{\textrm{ph}}=\omega/k$ (denoted by primed variables), the electromagnetic fields and the equation of motion transform to 
\begin{eqnarray}
\mathbf{B}^{\prime} = & & B_{0}\left[\hat{x}+b'\left(\hat{y}\sin\left(k^{\prime}x^{\prime}\right)+\hat{z}\cos\left(k^{\prime}x^{\prime}\right)\right)\right],\nonumber \\
\mathbf{E}^{\prime} = & & \mathbf{0},\label{eq: EM fields in wave frame}
\end{eqnarray}
\begin{equation}
\frac{\mathrm{d}}{\mathrm{d}t^{\prime}}\left(\frac{\mathbf{p}^{\prime}}{mc}\right) =-\omega_{\mathrm{c}}\frac{\mathbf{p}^{\prime}}{\gamma^{\prime}mc}\times\frac{\mathbf{B}^{\prime}}{B_{0}},\label{eq: e.o.m. in wave frame}
\end{equation}
where $\omega_{\textrm{c}}=\left|qB_{0}/m\right|$ is the lab-frame cyclotron frequency and all the primed symbol indicates wave-frame variables. Specifically, $b'=b\sqrt{1-n^{-2}}$ and $k^{\prime}=k\sqrt{1-n^{-2}}$, with the wave refractive index $n=ck/\omega=c/v_{\textrm{ph}}$. Therefore, in the wave frame, the electromagnetic fields reduce to a time-independent twisted magnetic field. Since $\mathbf{E}^{\prime}=0$, the particle energy in the wave frame is conserved; equivalently, the wave-frame particle Lorentz factor $\gamma^{\prime}$ is constant.

A dimensionless ``frequency mismatch parameter'' is now introduced as 
\begin{eqnarray}
\xi \equiv & & -\frac{\gamma}{\omega_{\textrm{c}}}\left(\omega-kv_{x}-\frac{\omega_{\textrm{c}}}{\gamma}\right),\nonumber\\
 = & & 1+\frac{ck^{\prime}}{\omega_{\textrm{c}}}\frac{p_{x}^{\prime}}{mc},\label{eq:frequency mismatch parameter}
\end{eqnarray}
which is zero at normal Doppler resonance $\omega-kv_{x}=\omega_{\textrm{c}}/\gamma$ and thus is a measure of how far the particle is away from this resonance. Because wave parameters are assumed to be constant, $\xi$ is a characteristic of the particle, embodying information about its velocity, energy, pitch-angle, etc, and so change in $\xi$ corresponds to changes in these particle parameters. Also, given the wave parameters, the resonant energy $\gamma_\textrm{r}$ and resonant parallel momentum $p_\textrm{r}$ can be calculated given the pitch-angle $\alpha$ by solving $\xi=0$. 

The vector equations (\ref{eq: EM fields in wave frame}) and (\ref{eq: e.o.m. in wave frame}) can be exactly recast into a scalar equation for $\xi$ \citep{Bellan2013}, which describes an equation of motion of the particle's $\xi$ in a pseudo-potential $\psi(\xi)$
\begin{equation}
\frac{\gamma^{\prime2}}{\omega_{\textrm{c}}^{2}}\frac{\mathrm{d}^{2}\xi}{\mathrm{d}t^{\prime2}}=-\frac{\partial\psi\left(\xi\right)}{\partial\xi},\label{eq:e.o.m. in terms of xi}
\end{equation}
where
\begin{equation}
\psi\left(\xi\right)=\frac{1}{8}\xi^{4}+\frac{1}{2}\left(b^{\prime2}-b^{\prime}\frac{ck^{\prime}}{\omega_{\textrm{c}}}\frac{p_{z0}^{\prime}}{mc}-\frac{1}{2}\xi_{0}^{2}\right)\xi^{2}-b^{\prime2}\xi,\label{eq:pseudo-potential}
\end{equation}
and subscript $0$ indicates initial conditions at $t=t'=0$, and the initial particle position $x\left(t=0\right)=0$ was assumed without loss of generality. (\ref{eq:pseudo-potential}) shows that the pseudo-potential is a quartic function of $\xi$, and its coefficients are determined by wave parameters and particle initial conditions.

From (\ref{eq:e.o.m. in terms of xi}), one can derive the conservation of pseudo-energy
\begin{equation}
\frac{1}{2}\left(\frac{\gamma^{\prime}}{\omega_{\textrm{c}}}\frac{\mathrm{d}\xi}{\mathrm{d}t^{\prime}}\right)^{2}+\psi\left(\xi\right)=\textrm{const.},\label{eq:pseudo-energy conservation}
\end{equation} which illustrates the conservative motion of the particle in $\left(\xi,\mathrm{d}\xi/\mathrm{d}t^{\prime}\right)$ space.

\section{Parallel electric field effects ($E_{0}\neq0$)}

Now, we add a finite steady-state electric field $\mathbf{E}=\mathbf{E}'=E_{0}\hat{x}$. To reduce notational clutter, we define normalized quantities $\bar{t}=\omega_{\textrm{c}}t$, $\bar{\mathbf{p}}=\mathbf{p}/mc$, $\bar{\mathbf{B}}=\mathbf{B}/B_{0},$ and $\bar{\mathbf{E}}=\mathbf{E}/cB_{0}$. (\ref{eq: e.o.m. in wave frame}) changes to
\begin{equation}
\frac{\mathrm{d}\bar{\mathbf{p}}'}{\mathrm{d}\bar{t}'}=-\bar{E}_{0}\hat{x}-\frac{\bar{\mathbf{p}}'}{\gamma^{\prime}}\times\left(\hat{x}+\mathbf{\bar{B}}_{\perp}^{\prime}\right).\label{eq:e.o.m. in wave frame, E0}
\end{equation}

With $E_{0}\neq0$, $\gamma^{\prime}$ is no longer constant, so particle motion in $\left(\xi,\mathrm{d}\xi/\mathrm{d}t^{\prime}\right)$ space is no longer conservative. Nevertheless, a new pseudo-energy equation can be derived exactly (detailed derivation in Appendix) and is
\begin{equation}
\frac{1}{2}\left(\gamma'\frac{\mathrm{d}\xi}{\mathrm{d}\bar{t}'}\right)^{2}+\Psi\left(\xi,\bar{t}'\right) \equiv W_{\textrm{tot}}=\textrm{const.},\label{eq:extended energy equation, E0}
\end{equation}
where the generalized pseudo-potential $\Psi\left(\xi,\bar{t}'\right)$ is
\begin{eqnarray}
\Psi\left(\xi,\bar{t}'\right) \equiv & & \frac{1}{8} \xi^{4}  + \frac{1}{2}  \left[b^{\prime2}-\bar{E}_{0}^{2}-b^{\prime}n_{\textrm{c}}'\bar{p}_{z0}^{\prime} \vphantom{n_{\textrm{c}}'\bar{E}_{0}\int_{0}^{\bar{t}'}\xi\left(\bar{\tau}^{\prime}\right)\mathrm{d}\bar{\tau}^{\prime}}   -  \frac{1}{2}\xi_{0}^{2} + n_{\textrm{c}}'\bar{E}_{0}\int_{0}^{\bar{t}'}\xi\left(\bar{\tau}^{\prime}\right)\mathrm{d}\bar{\tau}^{\prime}\right]\xi^{2} \nonumber\\
& & - \left(b^{\prime2} \right.  \left.- \bar{E}_{0}^{2}\right)\xi -\frac{1}{2}n_{\textrm{c}}'\bar{E}_{0}\int_{0}^{\bar{t}'}\xi^{3}\left(\bar{\tau}^{\prime}\right)\mathrm{d}\bar{\tau}^{\prime},\nonumber\\
& & \label{eq:Psi(xi,t')}
\end{eqnarray}
and $n_{\textrm{c}}'=ck'/\omega_{\textrm{c}}$. Note that $\Psi=\psi$ for $E_{0}=0$.  

Let us first examine particle dynamics far from resonance so that $\xi$ and $\xi_{0}$ are much larger than any other variable in (\ref{eq:Psi(xi,t')}). Dropping terms with $b'^{2}$, $b'$, and $\bar{E}_{0}^{2}$ in (\ref{eq:Psi(xi,t')}) and solving for $d\Psi/d\bar{t}'=0$ yields
\begin{equation}
\xi_{\textrm{far}}\simeq\xi_{0}-n_{\textrm{c}}'\bar{E}_{0}\bar{t}'.\label{eq:xi far resonance}
\end{equation}
The pseudo-kinetic-energy in (\ref{eq:extended energy equation, E0}) is then of order $\bar{E}_{0}^{2}$ and thus can be dropped. (\ref{eq:xi far resonance}) is the approximate solution for $\xi(\bar{t}')$ far away from $\xi=0$, and is the result of the particle being unaffected by the wave and only affected by $\bar{E}_{0}$ in (\ref{eq:e.o.m. in wave frame, E0}) so that $d\bar{p}_{x}'/d\bar{t}'=-\bar{E}_{0}$ or, equivalently, $d\xi/d\bar{t}'=-n_{\textrm{c}}'\bar{E}_{0}$.

As $\bar{t}'\rightarrow\xi_{0}/n_{\textrm{c}}'\bar{E}_{0}$, the particle approaches resonance ($\xi=0$). Because $\xi_{0}^{2}/2-n_{\textrm{c}}'\bar{E}_{0}\int_{0}^{\bar{t}'}\xi_{\textrm{far}}\left(\bar{\tau}'\right)d\bar{\tau}'\simeq\xi_{\textrm{far}}^{2}/2$, the quadratic coefficient in (\ref{eq:Psi(xi,t')}) becomes $\left(b^{\prime2}-b^{\prime}n_{\textrm{c}}'\bar{p}_{z0}^{\prime} \right.$ $\left.-\bar{E}_{0}^{2}\right)/2$ as $\xi\rightarrow0$. For a sufficiently large wave amplitude $b'$, this coefficient remains positive, ensuring that $\Psi$ traps the particle near $\xi=0$. If $b'$ is not sufficiently large, the coefficient turns negative, the particle is not trapped by the wave, and $\xi\rightarrow-\infty$ by (\ref{eq:xi far resonance}). Since the behavior of the integral terms becomes nonlinear near $\xi=0$, determining the exact trapping condition turns out to be rather intractable and is left for future work. Here, we focus on the behavior of the trapped particle.

Finally, we examine the dynamics of said resonance-trapped particle, whose $\xi$ oscillates around $\xi=0$. From the definition of $\xi$ in (\ref{eq:frequency mismatch parameter}), trapping implies that the particle $\bar{p}_{x}'$ oscillates around a constant value $\bar{p}_\textrm{r}'=-n_{\textrm{c}}'{}^{-1}$. Consequently, the bounce-averaged parallel momentum satisfies $\left\langle \bar{p}_{x}'\right\rangle =-n_{\textrm{c}}'{}^{-1}$ and $d\left\langle \bar{p}_{x}'\right\rangle /d\bar{t}'=0$. Dotting (\ref{eq:e.o.m. in wave frame, E0}) with $\bar{\mathbf{p}}'$, bounce-averaging, and finally using $d\left\langle \bar{p}_{x}'^2\right\rangle /d\bar{t}'=d\left\langle \bar{p}_{x}'\right\rangle^2 /d\bar{t}'$, we obtain solutions for the wave-frame momenta 
\begin{equation}
\left\langle \bar{p}_{\perp}'\right\rangle =\sqrt{\bar{p}'{}_{\perp\textrm{t}}^{2}+\frac{2\bar{E}_{0}}{n_{\textrm{c}}'}\left(\bar{t}'-\bar{t}_{\textrm{t}}'\right)}, \ \left\langle \bar{p}_{x}'\right\rangle =-\frac{1}{n_{\textrm{c}}'}\label{eq:average momentum in wave frame}
\end{equation}
where $\bar{t}_{\textrm{t}}'$ and $\bar{p}'_{\perp\textrm{t}}$ are the wave-frame time and perpendicular momentum at which the particle is resonance-trapped. (\ref{eq:average momentum in wave frame}) shows a surprising result that, once the particle is resonance-trapped, the parallel electric field $E_{0}$ energises the particle exclusively in the perpendicular direction and does not contribute to the average parallel momentum in the wave frame.

Now let us transform back to the lab frame. Because $\gamma'{}^{2}=1+\bar{p}'{}_{x}^{2}+\bar{p}'{}_{\perp}^{2}$, $d\left\langle \gamma'{}^{2}\right\rangle /d\bar{t}'=d\left\langle \bar{p}_{\perp}'^2\right\rangle /d\bar{t}'$ for a resonance-trapped particle and so $\left\langle \gamma'\right\rangle =\sqrt{\gamma'{}_{\textrm{t}}^{2}+2\bar{E}_{0}\left(\bar{t}'-\bar{t}_{\textrm{t}}'\right)/n_{\textrm{c}}'}$ where $\gamma'_{\textrm{t}}$ is the wave-frame Lorentz factor at trapping time. Using the Lorentz transformation $\bar{p}_{x}=\left(1-n^{-2}\right)^{-1/2}\left(\bar{p}_{x}'+\gamma'/n\right)$ and bounce-averaging, we obtain solutions for the lab-frame momenta
\begin{eqnarray}
\left\langle \bar{p}_{\perp}\right\rangle = &  & \left\langle \bar{p}_{\perp}^{\prime}\right\rangle \ ,\nonumber \\
\left\langle \bar{p}_{x}\right\rangle = &  & \frac{1}{\sqrt{1-n^{-2}}}\left(-\frac{1}{n_{\textrm{c}}'}+\frac{1}{n}\sqrt{\gamma_{\textrm{t}}'{}^{2}+\frac{2\bar{E}_{0}}{n_{\textrm{c}}'}\left(\bar{t}'-\bar{t}_{\textrm{t}}'\right)}\right), \label{eq:average momentum in lab frame}
\end{eqnarray}
which show another surprising result that once the particle is resonance-trapped, the particle is accelerated in the \textit{opposite} direction to the prior acceleration by $\bar{E}_{0}$, i.e., the particle is ``reflected'' from the resonant momentum $\bar{p}_{\textrm{r}}$ in parallel momentum space. 

Equations (\ref{eq:xi far resonance}), (\ref{eq:average momentum in wave frame}), and (\ref{eq:average momentum in lab frame}) are our main results, which can be summarized as follows. Under the presence of a uniform magnetic field, a transverse R-wave, and a parallel electric field, a negatively charged particle initially at far-resonance is accelerated by the electric field toward resonance without being affected much by the wave. Near resonance, the particle is trapped in resonance space if the wave amplitude is large enough. This trapped particle surprisingly experiences net acceleration in the perpendicular direction and also in the parallel direction opposite to its prior acceleration, i.e., it is reflected away from the resonant momentum. Therefore, the R-wave acts as a ``firewall'' in parallel momentum space, preventing further acceleration by the parallel electric field beyond the resonant momentum.

\begin{figure}
\centering
\includegraphics[width=0.85\columnwidth]{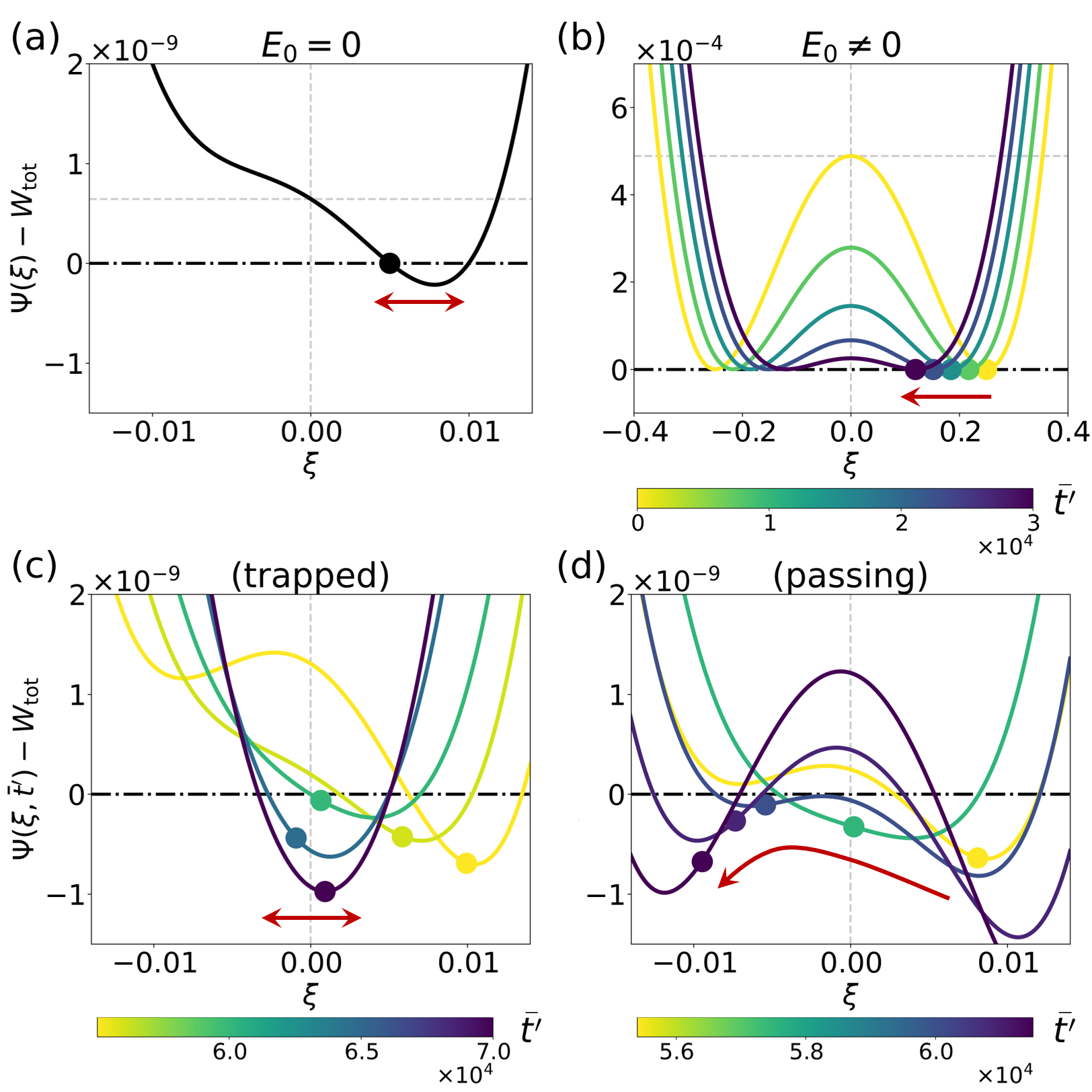}
\caption{\label{fig:Psi(xi,t')} Time evolution of $\Psi\left(\xi,\bar{t}^{\prime}\right)-W_{\textrm{tot}}$ for (a) $E_{0}=0$, (b) $\bar{E}_{0}=3.55\times10^{-5}$ and $b=5.0\times10^{-4}$, (c) same as (b), and (d) $\bar{E}_{0}=3.55\times10^{-5}$ and $b=3.5\times10^{-4}$. The dots are electron positions at each time, and the red arrows qualitatively describe electron motion.}
\end{figure}

\section{Single-particle simulation}

To verify and visualize the above predictions, we performed single-particle simulations using the relativistic Boris algorithm \citep{Birdsall2004}. Equation (\ref{eq:e.o.m. in wave frame, E0}) is solved for an electron in the wave frame with an electric field amplitude $\bar{E}_{0}=3.55\times10^{-5}$. In this configuration, all electrons are resonantly trapped for $b>4.75\times10^{-4}$, whereas all electrons pass through resonance for $b<3.9\times10^{-4}$. For intermediate amplitudes, trapping depends on the initial phase $\phi_{0} \equiv \sin^{-1} \left(\bar{p}_{z0}/\bar{p}_{\perp} \right)$. We consider two representative cases: $b=5.0\times10^{-4}$ (trapped) and $b=3.5\times10^{-4}$ (passing). The wave frequency and the wave number are set to $\bar{\omega}\equiv\omega/\omega_{\mathrm{c}}=0.152$ and $n_{\mathrm{c}}\equiv ck/\omega_{\mathrm{c}}=0.196$, consistent with the whistler wave dispersion relation for the resonant energy $\gamma_{\textrm{r}}=3$ and the normalized electron plasma frequency $\bar{\omega}_{\mathrm{p}}\equiv\sqrt{en_{0}/\varepsilon_{0}m}/\omega_{\mathrm{c}}=0.290$. These parameters correspond to $n=1.28$, $b'=3.16\times10^{-4}$ (trapped), $b'=2.20\times10^{-4}$ (passing), and $n'_\textrm{c}=0.123$. The initial momenta are chosen as $\bar{p}_{x0}^{\prime}=-6.10$ and $\bar{p}_{z0}^{\prime}=0.141$, corresponding to $\xi_{0}=0.25$. For this setup, the resonant momentum is $\bar{p}_\textrm{r}'=-n_\textrm{c}'^{-1}=-8.13$ in the wave frame and $\bar{p}_\textrm{r}=-2.79$ in the lab frame with $\alpha=0$. A reference simulation with $E_{0}=0$ and different particle parameters is also performed for comparison.

Figure \ref{fig:Psi(xi,t')} illustrates the time evolution of the generalized pseudo-potential $\Psi\left(\xi,\bar{t}^{\prime}\right)-W_{\textrm{tot}}$. Each point in the figure represents the electron's position in $\Psi-\xi$ space. For the reference $E_{0}=0$ case (Figure \ref{fig:Psi(xi,t')}(a)), $\Psi\left(\xi,\bar{t}^{\prime}\right)$ reduces to $\psi\left(\xi\right)$; its shape remains constant and is determined solely by the initial conditions, and the electron oscillates in $\xi$-space in this $\psi$. Figure \ref{fig:Psi(xi,t')}(b) and (c) show the evolution of $\Psi$ for $\bar{E}_{0}=3.55\times10^{-5}$ and $b=5.0\times10^{-4}$ case. As predicted by Eqs. (\ref{eq:Psi(xi,t')}-\ref{eq:xi far resonance}), the magnitude of the negative quadratic coefficient decreases with time (Figure \ref{fig:Psi(xi,t')}(b)). Near resonance (Figure \ref{fig:Psi(xi,t')}(c)), the quadratic coefficient of $\Psi$ becomes positive, resulting in a single-well potential that traps the electron around $\xi=0$. In contrast, for a passing electron with $\bar{E}_{0}=3.55\times10^{-5}$ and $b=3.5\times10^{-4}$ (Figure \ref{fig:Psi(xi,t')}(d)), the quadratic coefficient returns to a negative value, and the electron transits to the left well of the pseudo-potential.

\begin{figure}
\centering
\includegraphics[width=0.85\columnwidth]{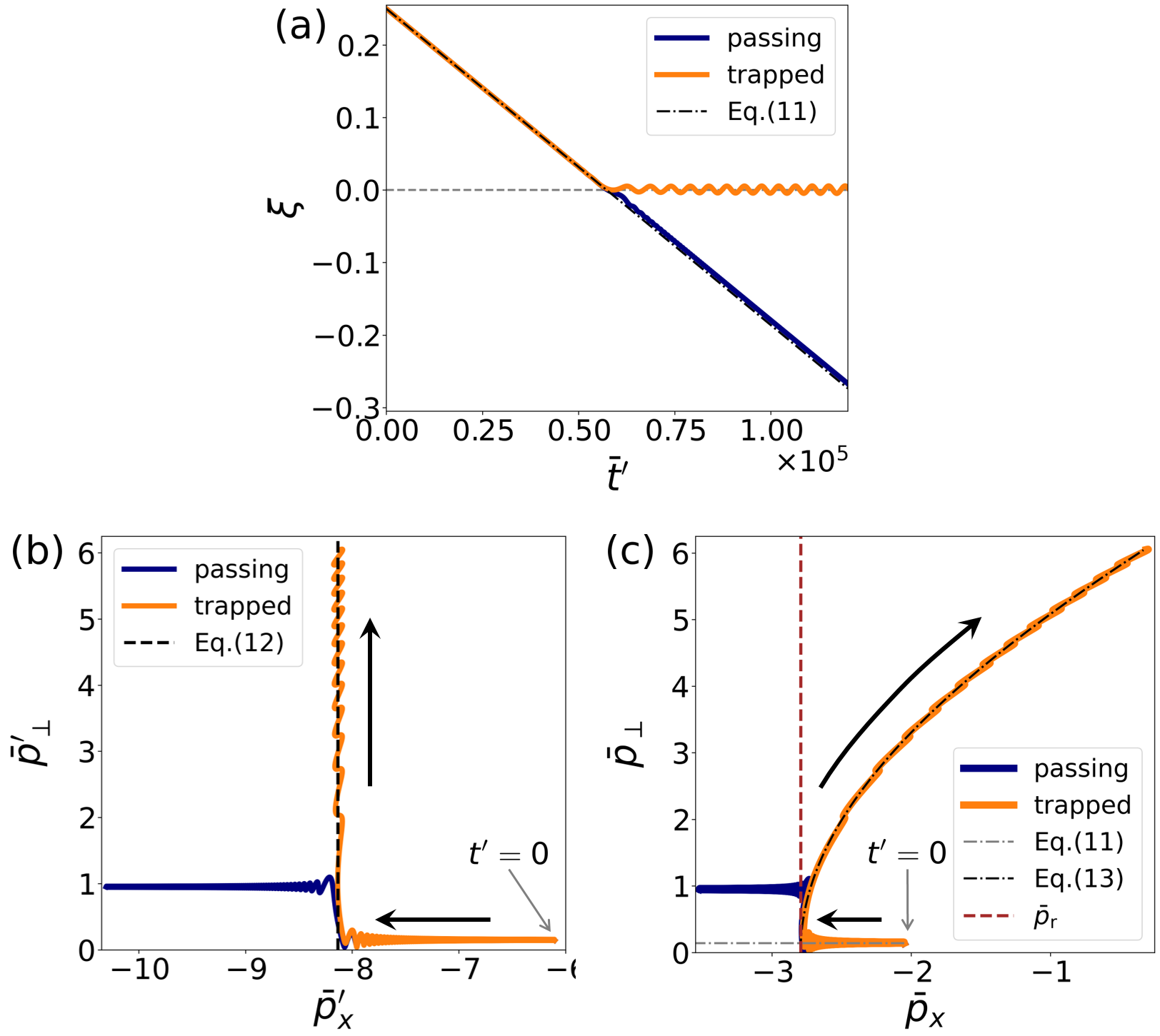}
\caption{\label{fig:xi+dxi_dt}(a) Time evolution of $\xi$, (b) electron trajectory in $\bar{p}_{\perp}^{\prime}-\bar{p}_{x}^{\prime}$ space, and (c) electron motion in $\bar{p}_{\perp}-\bar{p}_{x}$ space. The red vertical dashed line in (c) indicates the resonant momentum $\bar{p}_{\textrm{r}}$ with $\alpha=0$. The black arrows indicate the direction of electron trajectory. Note that $p_{x}$ changes opposite to the electrostatic force in the lab frame.}
\end{figure}

Figure \ref{fig:xi+dxi_dt}(a) shows the time-dependent electron trajectory in $\xi$-space. Consistent with Figure \ref{fig:Psi(xi,t')}(c), the trapped electron oscillates around $\xi=0$. The passing electron, which continues toward $\xi\rightarrow -\infty$, is shown for reference. In $\bar{p}_{\perp}^{\prime}-\bar{p}_{x}^{\prime}$ space (Figure \ref{fig:xi+dxi_dt}(b)), $\left\langle\bar{p}_x'\right\rangle$ is fixed to $-n_\textrm{c}'^{-1}$ as predicted by (\ref{eq:average momentum in wave frame}), and only $\left\langle\bar{p}_{\perp}^{\prime}\right\rangle$ increases in time. Figure \ref{fig:xi+dxi_dt}(c) shows the electron trajectory in the lab-frame $\bar{p}_{\perp}-\bar{p}_{x}$ space. As predicted in (\ref{eq:average momentum in lab frame}), $\bar{p}_{x}$ reverses direction, opposite to the electric force, while $\bar{p}_{\perp}$ continues to increase. These electron trajectories in full three-dimensional momentum space correspond to those in Figure \ref{fig:schematics}, albeit sparse-sampled for better presentation.

\section{Particle-in-cell Simulation}

Let us now check whether the single particle mechanism is robust with respect to self-consistent particle-in-cell (PIC) simulations.  We verify the firewall effect with an open-source PIC code SMILEI \citep{Derouillat2018}, in the context of runaway electrons in tokamak plasmas with a background magnetic field $B_{0}=3.5\ \left[\mathrm{T}\right]$, electron density $n_{0}=10^{19}\ \left[\mathrm{m^{-3}}\right]$, and bulk electron temperature $T_{\mathrm{e}0}=1\ \left[\mathrm{keV}\right]$. A 2D slab geometry with periodic boundary conditions is assumed, and the field profile modeling an externally-injected wave is prescribed as (\ref{eq:EM fields in lab frame}), with the same wave parameters as the single-particle simulation. To avoid numerical instabilities arising from uniform electric field acceleration, we benchmark a recent study modeling wave generation by runaway electrons \citep{Kang2024} where the plasma consists of immobile ions and electrons with a bump-on-tail momentum distribution; the latter undergoes an instability and forms a tail distribution, similar to the actual runaway distribution \citep{Paz-Soldan2017,Lvovskiy2018,Lvovskiy2019}. The initial bump population density is set to $n_{0}/16$, with average momenta $\bar{p}_{x0,\textrm{bump}}=-1.11$, and $\bar{p}_{\perp0,\textrm{bump}}=0.11$. All species are assumed collisionless and uniformly distributed in configuration space. We simulate two cases: (i) with an applied wave ($b=5\times10^{-4}$) and (ii) without a wave ($b=0$). Further details of the simulation setup are provided in the Appendix.

\begin{figure}
\includegraphics[width=1\columnwidth]{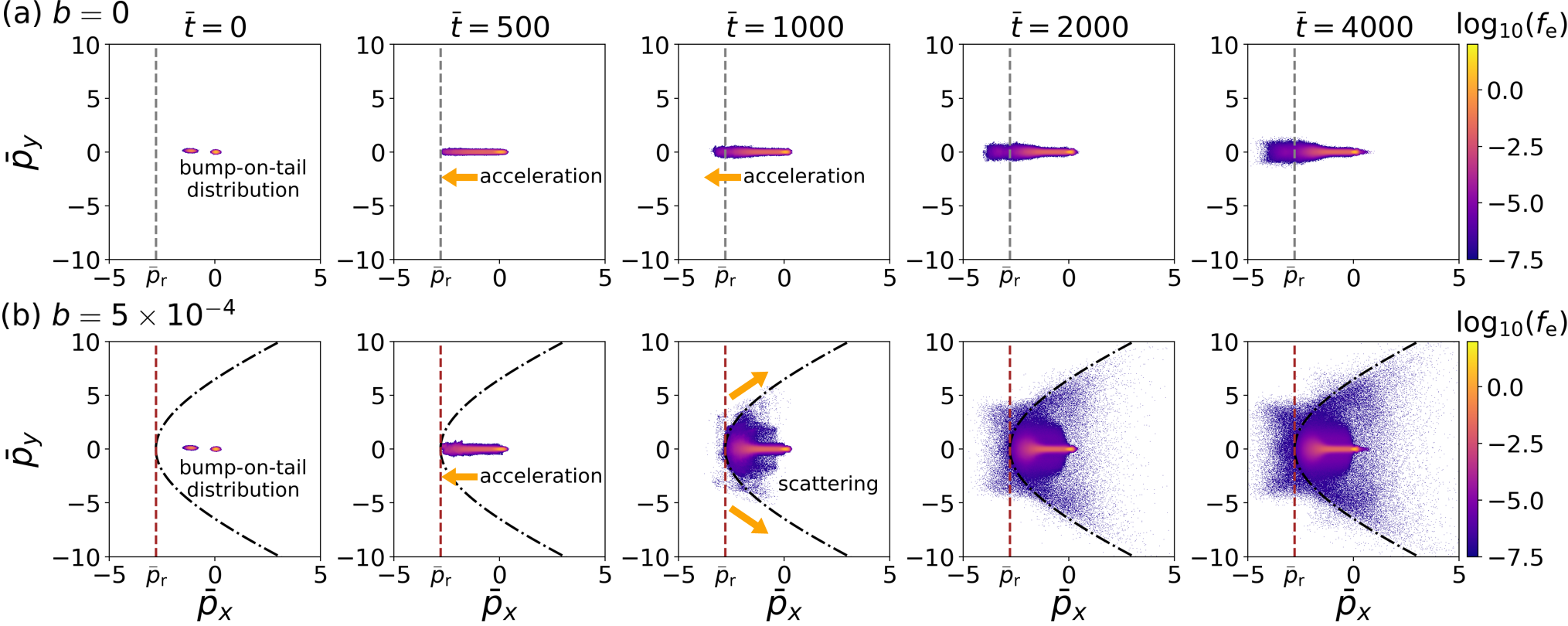}
\caption{\label{fig:fe(px,py,t)}Snapshots of electron momentum distribution $f_{\mathrm{e}}\left(\bar{p}_{x},\bar{p}_{y},\bar{t}\right)$ from the particle-in-cell simulation for the case (a) without external wave and (b) with wave. The black dash-dotted lines in (b) correspond to (\ref{eq:average momentum in lab frame}) substituting $\bar{p}'_{\perp\textrm{t}}=\bar{p}_{\perp0,\textrm{bump}}$ and $\bar{t}'_{\textrm{t}}=\xi_{0}/n'_{\textrm{c}}\bar{E}_{0}$, with $\xi_{0} = 1+n'_{\textrm{c}} \ \bar{p}_{x0,\textrm{bump}}$ and $\bar{E}_{0}=\left\langle\bar{E}_{x \ \textrm{rms}} \right\rangle\simeq8.11\times10^{-3}$, and gray or red vertical dashed lines in snapshots refer to $\bar{p}_{\textrm{r}}$ with $\alpha=0$.}
\end{figure}

Figure \ref{fig:fe(px,py,t)} shows snapshots of the electron distribution function $f_{\mathrm{e}}\left(\bar{p}_{x},\bar{p}_{y},\bar{t}\right)$---note that $\bar{p}_y$ is a proxy for $\bar{p}_\perp$. During $\bar{t}=0\sim500$, electrons are accelerated toward the target momentum in both cases as the bump-on-tail instability progresses. For $\bar{t}>500$, the case without wave has a significant population of electrons that are accelerated beyond $\bar{p}_\textrm{r}$. However, for the case with wave, most electrons are reflected from  $\bar{p}_\textrm{r}$ and experience enhanced pitch-angle scattering. Comparing the two cases, it is clear that the wave suppresses parallel acceleration beyond the resonant momentum via prompt scattering and back-acceleration. In Figure \ref{fig:fe(px,py,t)}(b), electrons are scattered along the black dash-dotted line, which corresponds to (\ref{eq:average momentum in lab frame}) substituting $\bar{p}'_{\perp\textrm{t}}=\bar{p}_{\perp0,\textrm{bump}}$ and $\bar{t}'_{\textrm{t}}=\xi_{0}/n'_{\textrm{c}}\bar{E}_{0}$ (from (\ref{eq:xi far resonance})), with $\xi_{0} = 1+n'_{\textrm{c}} \ \bar{p}_{x0,\textrm{bump}}$ and $\bar{E}_{0}=\left\langle\bar{E}_{x \ \textrm{rms}} \right\rangle\simeq8.11\times10^{-3}$, where $\bar{E}_{x,\textrm{rms}} =\left[\bar{L}_{x}^{-1}\bar{L}_{y}^{-1}\int_{0}^{\bar{L}_{y}}\mathrm{d}\bar{y}\int_{0}^{\bar{L}_{x}}\mathrm{d}\bar{x}  \left( \bar{E}_{x}^{2} \right) \right]^{1/2}$. This confirms that the scattering mechanism observed in the simulation aligns with the single-particle theoretical prediction.

Quantitatively, the energetic electron (EE) population density with $\left|\bar{p}_{x}\right|>\left|\bar{p}_{\textrm{r}}\right|$, $n_{\mathrm{EE}}\equiv\int_{0}^{\infty}\mathrm{d}\bar{p}_{\perp}\int_{-\infty}^{\bar{p}_{\textrm{r}}}\mathrm{d}\bar{p}_{x}\left(f_{\mathrm{e}}\right)$, is reduced by 87\%, and the RMS value of $\bar{p}_{y}$, $\bar{p}_{y,\mathrm{EE},\textrm{rms}} \equiv\left[ n_{\mathrm{EE}}^{-1}\int_{0}^{\infty}\mathrm{d}\bar{p}_{\perp}\int_{-\infty}^{\bar{p}_{\textrm{r}}}\mathrm{d}\bar{p}_{x}\left(\bar{p}_{y}^{2}f_{\mathrm{e}}\right)\right]^{1/2}$ increases by about 516\%, compared to the case without wave. Therefore, an externally-injected wave can dramatically limit the momenta of runaways and scatter their pitch-angles.

\section{Discussion and Summary}

Quasilinear diffusion is frequently invoked to describe the evolution of runaway distribution in tokamaks \citep{Pokol2008, Liu2018, Guo2018}. Although quasilinear theory does not apply for coherent waves, let us nonetheless compare the pitch-angle-scattering timescale inferred from the simulation results with that predicted by quasilinear diffusion theory.  Using the saturation time of $n_{\mathrm{EE}}$ and $\left\langle \bar{p}_{y}^{2}\right\rangle _{\mathrm{EE}}$, we estimate a characteristic pitch-angle scattering time of $\bar{\tau}_{\alpha,\textrm{sim}}\sim10^{3}$. On the other hand, applying the quasilinear diffusion coefficient of pitch angle scattering $D_{\alpha\alpha}$ using Eq. (36) in \citep{Summers2005}, evaluated for a Gaussian centered at $\bar{\omega}=0.152$ with the standard deviation $\Delta\bar{\omega}=0.05$, yields $\bar{D}_{\alpha\alpha}\sim5\times10^{-10}$, and $\bar{\tau}_{\alpha,\textrm{QL}}\sim1/D_{\alpha\alpha}\simeq2\times10^{9}$. Thus, pitch-angle scattering induced by the coherent R-wave proceeds roughly $10^{6}$ times faster than expected from quasilinear diffusion, hinting the potential for an extremely rapid runaway suppression method.

We next estimate the power required to compensate for the collisional damping of the wave inside a tokamak \citep{Guo2018,Yoon2021,Aleynikov2015}. For representative tokamak disruption parameters ($B_{0}=3.5\ \left[\mathrm{T}\right]$, $n_{0}=10^{19}\ \left[\mathrm{m^{-3}}\right]$, $T_{\mathrm{e}0}=100\ \left[\mathrm{eV}\right]$, $Z=1$), the electron-ion collision frequency and R-wave damping rate are $\nu_{\mathrm{ei}}\simeq0.428\ \left[\mathrm{MHz}\right]$, $\left|\omega_{i}\right|/2\pi\simeq7.13\ \left[\mathrm{kHz}\right]$, respectively. Using the power balance equation $P=\left(\left|\omega_{i}\right|/2\pi\right)\left(b^{2}B_{0}^{2}/2\mu_{0}\right)V_{\mathrm{p}}$ and the KSTAR plasma volume $V_{\mathrm{p}}=17.8\ \left[\mathrm{m^{3}}\right]$, we obtain required power of approximately $P\simeq155\ \left[\mathrm{kW}\right]$. Because of the rapid interaction, the applied wave only needs to fill a small part of the volume,  and so this value is a vast overestimation and thus represents an upper bound. However, the injection efficiency of R-waves or whistler waves to the tokamak core remains an open question, as one may require complicated methods such as mode conversion to achieve this feat. 

It should be noted that although a constant electric field was assumed in the theoretical analysis, the parallel field in the PIC simulation is a wave field arising from bump-on-tail instability (snapshots of the electric field are shown in the Appendix). The fact that the same firewall effect was observed in the latter means that the mechanism is robust to small changes in the spatiotemporal nature of the parallel electric field.

For further fusion contexts, if (\ref{eq: EM fields in wave frame}) is assumed to be in the lab frame, it simply describes a current-free twisted magnetic field profile akin to the local field profile in stellarators, with $k'$ encapsulating information about the rotational transform. Therefore, although the present setup is much simpler, the firewall effect is likely present in stellarators as well. This means that, given appropriate conditions, the presence of a parallel electric field in stellarators may induce perpendicular energization and perhaps deconfinement, and further investigation is thus warranted. In mirror devices, the present mechanism may be leveraged to increase the pitch-angles of particles and push them away from the loss cone.

In space, it is well-known that energetic particles in the magnetosphere can resonantly interact with coherent circularly-polarized waves and undergo significant pitch-angle scattering \citep{Bellan2013,Yoon2020}. In previous studies, parallel electric fields would naturally have been thought of as a mechanism to decrease the pitch-angles of the particles. As we clearly show here, parallel electric fields in fact increase the pitch-angles of resonantly trapped particles, and so the present firewall effect warrants a revisiting of the particle scattering theories in space contexts.

In summary, we have identified a counterintuitive firewall effect in which a coherent right-handed circularly polarized wave, combined with a constant parallel electric field, reverses the parallel acceleration direction of an electron and also accelerates it in the perpendicular direction, once it is trapped in Doppler-shifted cyclotron resonance. Single-particle simulations and self-consistent particle-in-cell simulations corroborate this mechanism, demonstrating its ability to suppress further runaway-electron acceleration in fusion-relevant conditions. More broadly, the underlying mechanism provides a novel perspective on resonant pitch-angle scattering in magnetized plasmas—from tokamaks and stellarators to astrophysical and space environments.

\section*{Acknowledgments}
The computational resources for the simulations were provided by the KAIROS supercomputing system at the Korea Institute of Fusion Energy (KFE). This work was supported by the National Research Foundation (NRF) of Korea, grant Nos. RS-2022-00154676, RS-2023-00281272, RS-2024-00409564, (G.S.Y. and H.L.K.), RS-2025-00522068 (Y. D. Y.), and RS-2024-00455499 (M. H. C.). Y.D.Y. was supported by an appointment to the JRG program at the APCTP through the Science and Technology Promotion Fund of the Korean Government. Y.D.Y. was also supported by the Korean Local Governments — Gyeongsangbuk-do Province and Pohang City.


\appendix

\setcounter{equation}{0}
\setcounter{figure}{0}
\renewcommand{\theequation}{A\arabic{equation}} 
\renewcommand{\thefigure}{A\arabic{figure}}

\section{Derivation of the pseudo-energy equation}

Here, we derive Eqs. (\ref{eq:extended energy equation, E0}-\ref{eq:Psi(xi,t')}) in the main text. Separating the equation of motion in the wave frame ((\ref{eq: e.o.m. in wave frame})) into the parallel and perpendicular directions, we obtain

\begin{eqnarray}
\frac{\mathrm{d}\bar{p}'_{x}}{\mathrm{d}\bar{t}'}= & & -\bar{E}_{0}- \hat{x} \cdot \frac{\bar{\mathbf{p}}'_{\perp}}{\gamma^{\prime}}\times \mathbf{\bar{B}}_{\perp}^{\prime},\label{eq: dp'_x/dt'} \\
\frac{\mathrm{d}\bar{\mathbf{p}}'_{\perp}}{\mathrm{d}\bar{t}'}= & & -\frac{\bar{\mathbf{p}}'_{\perp}}{\gamma^{\prime}}\times \hat{x} -\frac{\bar{p}'_{x}}{\gamma^{\prime}} \hat{x} \times \mathbf{\bar{B}}_{\perp}^{\prime}.\label{eq: dp'_perp/dt'}
\end{eqnarray}
The time derivative of (\ref{eq: dp'_x/dt'}) is

\begin{eqnarray}
\frac{\mathrm{d}^{2}\bar{p}'_{x}}{\mathrm{d}\bar{t}'^{2}} = &&-\hat{x}\cdot\left[\left\{ \mathbf{\bar{p}}_{\perp}^{\prime} \frac{\mathrm{d} }{\mathrm{d}\bar{t}^{\prime}} \left( \frac{1}{\gamma^{\prime}} \right) +\frac{1}{\gamma^{\prime}} \frac{\mathrm{d} \mathbf{\bar{p}}'_{\perp}}{\mathrm{d}\bar{t}^{\prime}} \right\}\times\mathbf{\bar{B}}_{\perp}^{\prime} + \frac{\mathbf{\bar{p}}_{\perp}^{\prime}}{\gamma^{\prime}}\times\frac{\mathrm{d} \mathbf{\bar{B}}_{\perp}^{\prime}}{\mathrm{d}\bar{t}^{\prime}} \right].\label{eq: second derivative of px'}
\end{eqnarray}
Using (\ref{eq: dp'_perp/dt'}) and the relations below,

\begin{align}
\frac{1}{2}\frac{\mathrm{d} \bar{p}'^{2}}{\mathrm{d}\bar{t}^{\prime}} &= \mathbf{\bar{p}'}\cdot \frac{\mathrm{d} \mathbf{\bar{p}}'}{\mathrm{d}\bar{t}^{\prime}} = -\bar{E}_{0}\bar{p}'_{x},\nonumber\\
\frac{\mathrm{d}}{\mathrm{d}\bar{t}^{\prime}}\left(\frac{1}{\gamma^{\prime}}\right) &= -\frac{1}{2\gamma'^{3}} \frac{\mathrm{d} \bar{p}'^{2}}{\mathrm{d}\bar{t}^{\prime}}=\frac{\bar{E}_0\bar{p}_x'}{\gamma'^3},\nonumber\\
\hat{x} \cdot \frac{\bar{\mathbf{p}}'_{\perp}}{\gamma'} \times \mathbf{\bar{B}}_{\perp}^{\prime} &= - \left( \bar{E}_{0} + \frac{\mathrm{d}\bar{p}'_{x}}{\mathrm{d}\bar{t}'} \right),\nonumber
\end{align}
and
\begin{eqnarray*}
\frac{\mathrm{d} \mathbf{\bar{B}}_{\perp}^{\prime}}{\mathrm{d}\bar{t}^{\prime}} = & & \frac{\mathrm{d} }{\mathrm{d}\bar{t}^{\prime}} b'\left(\hat{y}\sin\left(k^{\prime}x^{\prime}\right)+\hat{z}\cos\left(k^{\prime}x^{\prime}\right)\right) \\
= & & -n'_{\textrm{c}} \frac{\bar{p}'_{x}}{\gamma^{\prime}} \hat{x}\times \mathbf{\bar{B}}'_{\perp},
\end{eqnarray*}
(\ref{eq: second derivative of px'}) becomes

\begin{eqnarray}
\frac{\mathrm{d}^{2}\bar{p}'_{x}}{\mathrm{d}\bar{t}'^{2}} = & & \frac{1}{\gamma^{\prime2}} \left[  \bar{E}_{0}\bar{p}'_{x} \left( \bar{E}_{0} + \frac{\mathrm{d}\bar{p}'_{x}}{\mathrm{d}\bar{t}'} \right) + \left(1 + n'_{\textrm{c}} \bar{p}'_{x} \right) \left(\mathbf{\bar{p}}_{\perp}^{\prime}\cdot \mathbf{\bar{B}}_{\perp}^{\prime} \right) - \bar{p}_{x}^{\prime}b'^{2}  \right].
\label{eq: second derivative of px' 2}
\end{eqnarray}
At this point, we utilize the definition of the frequency mismatch parameter in (\ref{eq:frequency mismatch parameter}) and also note that 
\begin{align}
    \frac{\mathrm{d}\bar{p}_x'}{\mathrm{d}\bar{t}'}=\frac{1}{n_\mathrm{c}'}\frac{\mathrm{d}\xi}{\mathrm{d}\bar{t}'}.
\end{align}
From the time derivative of $\mathbf{\bar{p}}'_{\perp} \cdot \mathbf{\bar{B}}'_{\perp}$, we obtain

\begin{eqnarray}
\frac{\mathrm{d}}{\mathrm{d}\bar{t}'} \left( \mathbf{\bar{p}}'_{\perp}\cdot \mathbf{\bar{B}}'_{\perp}\right)
= & & \frac{\mathrm{d} \mathbf{\bar{p}}'_{\perp}}{\mathrm{d}\bar{t}'} \cdot \mathbf{\bar{B}}'_{\perp}+ \mathbf{\bar{p}}'_{\perp}\cdot\frac{\mathrm{d}\mathbf{\bar{B}}'_{\perp}}{\mathrm{d}\bar{t}'},\nonumber\\
= & & \left(1+n'_{\textrm{c}}\bar{p}'_{x} \right) \left(\hat{x} \cdot \frac{\mathbf{\bar{p}}'_{\perp}}{\gamma^{\prime}} \times \mathbf{\bar{B}}'_{\perp} \right)\nonumber,\\
= & & - \xi\left( \bar{E}_{0} + \frac{\mathrm{d}\bar{p}'_{x}}{\mathrm{d}\bar{t}'} \right)\nonumber,\\
= & & -  \bar{E}_{0}\xi- \frac{1}{2n'_{\textrm{c}}} \frac{\mathrm{d}\xi^2}{\mathrm{d}\bar{t}'} .\label{eq: d(p'_perp * B'_perp)/dt'}
\end{eqnarray}
Integrating over time and using $\left. \mathbf{\bar{p}}'_{\perp} \cdot \mathbf{\bar{B}}'_{\perp}\right|_{\bar{t}'=0}=b'\bar{p}_{z0}'$,
\begin{eqnarray}
\mathbf{\bar{p}}'_{\perp} \cdot \mathbf{\bar{B}}'_{\perp} = & & b'\bar{p}_{z0}' - \bar{E}_{0} \int_{0}^{\bar{t}'} \xi\left( \bar{\tau}'\right) \mathrm{d}\bar{\tau}'  - \frac{1}{2n'_{\textrm{c}}} \left(\xi^{2} -\xi_0^{2}\right).
\end{eqnarray}
Using the above and replacing all $\bar{p}_x'$ by $n_\textrm{c}'^{-1}\left(\xi-1\right)$, (\ref{eq: second derivative of px' 2}) becomes
\begin{eqnarray}
\gamma^{\prime2} \frac{\mathrm{d}^{2}\xi}{\mathrm{d}\bar{t}^{\prime2}} & =& \left(\xi-1\right)\bar{E}_{0}\left(\bar{E}_{0}+\frac{1}{n'_{\textrm{c}}}\frac{\mathrm{d}\xi}{\mathrm{d}\bar{t}^{\prime}}\right) - \left( \xi  -1 \right)b'^{2} \nonumber\\
& & + \xi \left[b'n'_{\textrm{c}}\bar{p}'_{z0} - n'_{\textrm{c}}\bar{E}_{0} \int_{0}^{\bar{t}'} \xi\left( \bar{\tau}'\right) \ \mathrm{d}\bar{\tau}'  - \frac{1}{2} \left( \xi^{2} -\xi^{2}_{0} \right) \right].
\label{eq: second derivative of xi}
\end{eqnarray}
Using 

\begin{eqnarray*}
\frac{1}{2}\frac{\mathrm{d}\gamma^{\prime2}}{\mathrm{d}\bar{t}^{\prime}} = \frac{1}{2}\frac{\mathrm{d}\bar{p}_x'^2}{\mathrm{d}\bar{t}'}= & & -\bar{E}_{0} \bar{p}_{x}^{\prime}\\
= & & -\frac{\bar{E}_{0}}{n'_{\textrm{c}}}\left(\xi-1\right),
\end{eqnarray*}
and

\begin{equation*}
\xi\int_{0}^{\bar{t}^{\prime}}\xi\left(\bar{\tau}^{\prime}\right)\mathrm{d}\bar{\tau}^{\prime} = \frac{1}{2}\frac{\partial}{\partial\xi}\left(\xi^{2}\int_{0}^{\bar{t}^{\prime}}\xi\left(\bar{\tau}^{\prime}\right)\mathrm{d}\bar{\tau}^{\prime}-\int_{0}^{\bar{t}^{\prime}}\xi^{3}\left(\bar{\tau}^{\prime}\right)\mathrm{d}\bar{\tau}^{\prime}\right),
\end{equation*}
we obtain

\begin{equation}
\gamma^{\prime2} \frac{\mathrm{d}^{2}\xi}{\mathrm{d}\bar{t}^{\prime2}} = -\frac{1}{2}\frac{\mathrm{d}\gamma^{\prime2}}{\mathrm{d}\bar{t}^{\prime}}\frac{\mathrm{d}\xi}{\mathrm{d}\bar{t}^{\prime}} - \frac{\partial}{\partial \xi} \Psi \left(\xi,\bar{t}'\right),\label{eq: second derivative of xi (final)}
\end{equation}
where the generalized pseudo-potential $\Psi\left(\xi,\bar{t}'\right)$ is described in (\ref{eq:Psi(xi,t')}).
Multiplying $\mathrm{d}\xi/\mathrm{d}\bar{t}'$ to (\ref{eq: second derivative of xi (final)}),  and integrating over $\bar{t}'$, we finally obtain the pseudo-energy equation: (\ref{eq:extended energy equation, E0}).

\setcounter{equation}{0}
\setcounter{figure}{0}
\setcounter{table}{0}
\renewcommand{\theequation}{B\arabic{equation}} 
\renewcommand{\thefigure}{B\arabic{figure}}
\renewcommand{\thetable}{B\arabic{table}}

\section{Details of PIC simulation}
In this section, we present the PIC simulation setup and the time evolution of the electric fields in more detail. The simulation code and geometry are described in the main text. In addition, we employ the Lehe field solver \citep{Lehe2013} to mitigate numerical Cherenkov instability. The detailed simulation parameters are summarized in Table \ref{tab:PIC_sim_param}.

\begin{table}
\begin{center}
\begin{tabular}{lll}

Wave frequency & & \ $\bar{\omega} = 0.152$  \\
Wave number &  &  \ $n_{\textrm{c}} = 0.196$  \\
Number of cells &  & \ $N_{\textrm{cell},x}=2^{14}$,\\
&  & \ $N_{\textrm{cell},y}=2^{9}$\\
Domain size &  & \ $\bar{L}_{x}=5 \ \bar{\lambda}$,\\
&  & \ $\bar{L}_{y}=1.25 \ \bar{\lambda}$\\
Timestep &  & \ $\Delta\bar{t}=9.20\times10^{-3}$\\
Simulation time &  & \ $\bar{t}_{\textrm{sim}}=5000$\\
Electron plasma frequency &  & \ $\bar{\omega}_{\mathrm{p}} = 0.290$ \\
Initial average momentum &  & \ $\bar{p}_{x0,\textrm{BE}}=0.0495$,\\
(background electron) &  & \ $\bar{p}_{\perp0,\textrm{BE}}=0$\\
Initial average momentum &  & \ $\bar{p}_{x0,\textrm{bump}}=-1.11$,\\
(bump) &  & \ $\bar{p}_{\perp0,\textrm{bump}}=0.11$\\
Initial bump population &  & \ $n_{0,\textrm{bump}}=n_{0}/16$ \\
Temperature &  & \ $T_{0,\textrm{BE}}=T_{0,\textrm{bump}}=1 \ \left[ \mathrm{keV} \right]$ \\
Super-particles per cell &  & \ $N_{\textrm{ppc}}=\begin{cases}
64, & \text{for BE}\\
32, & \text{for ion, bump}
\end{cases}$\\

\end{tabular}

\caption{\label{tab:PIC_sim_param} PIC simulation parameters. Here, $\bar{\lambda}\equiv2\pi/n_{\textrm{c}}$.}

\end{center}
\end{table}

\begin{figure}
\centering
\includegraphics[width=0.85\columnwidth]{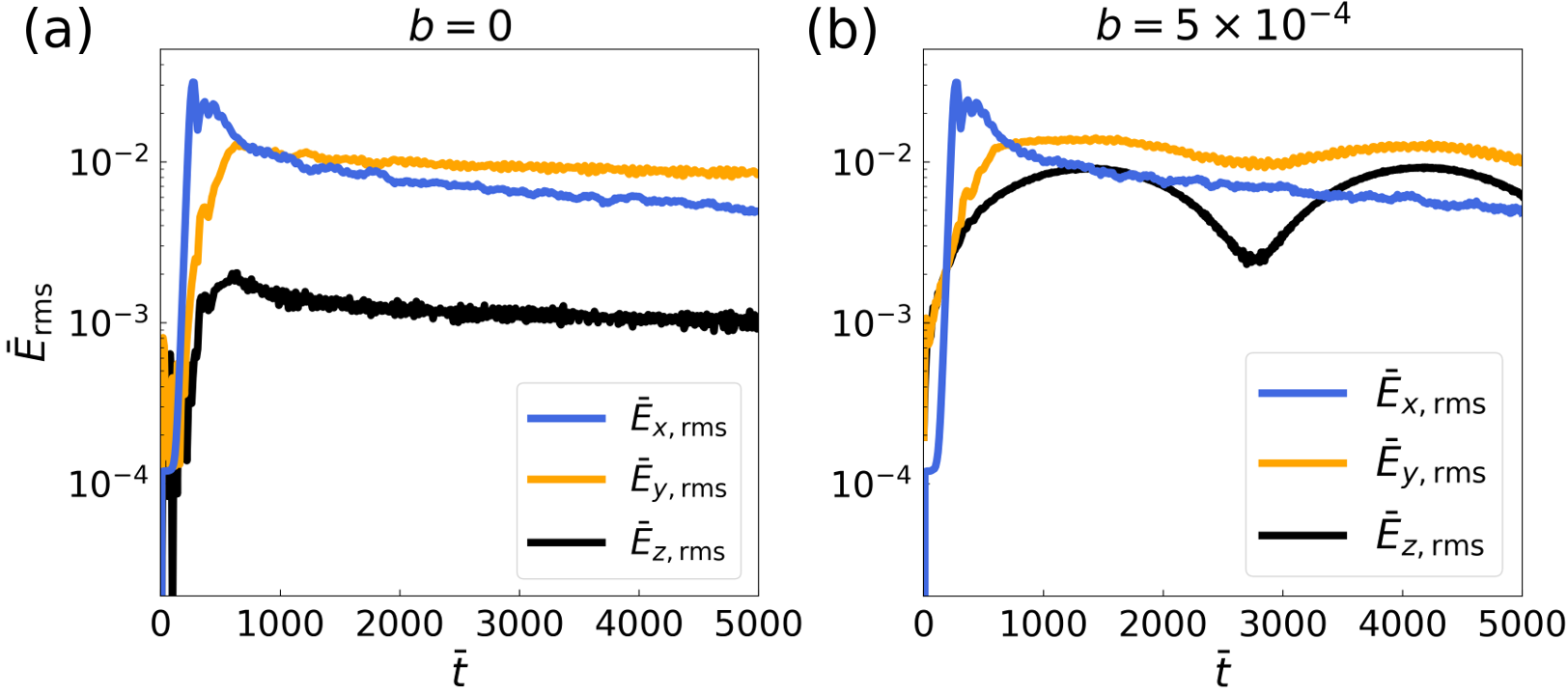}
\caption{\label{fig:E_rms} Log scale plot of $\bar{E}_{j,\mathrm{rms}}$ ($j=x,y,z$) for (a) $b=0$ and (b) $b=5\times10^{-4}$.}
\end{figure}

Figure \ref{fig:E_rms} shows the evolution of the RMS electric fields $\bar{E}_{j,\mathrm{rms}}$ ($j=x,y,z$) for (a) $b=0$ and (b) $b=5\times10^{-4}$. For $\bar{t}<250$, an electrostatic wave is driven by the relaxation of the bump-on-tail distribution in the $x$ direction \citep{Kang2024}. By Maxwell’s equations in the 2D geometry ($\Delta z=0$), $\bar{B}_{z}$ and $\bar{E}_{y}$ grow predominantly in time for the $b=0$ case. In contrast, as shown in Figure \ref{fig:E_rms}(b), resonant interactions between the R-wave and electrons enhance both $\bar{E}_{y}$ and $\bar{E}_{z}$. For $\bar{t}>1000$, the electric fields saturate at the level $\bar{E}_{\textrm{rms}}\sim10^{-3}$. The time-averaged values of $\bar{E}_{x,\textrm{rms}}$ are $\langle\bar{E}_{x,\textrm{rms}}\rangle\simeq8.04\times10^{-3}$ for $b=0$ and $\langle\bar{E}_{x,\textrm{rms}}\rangle\simeq8.11\times10^{-3}$ for $b=5\times10^{-4}$.

\begin{figure}
\includegraphics[width=1\columnwidth]{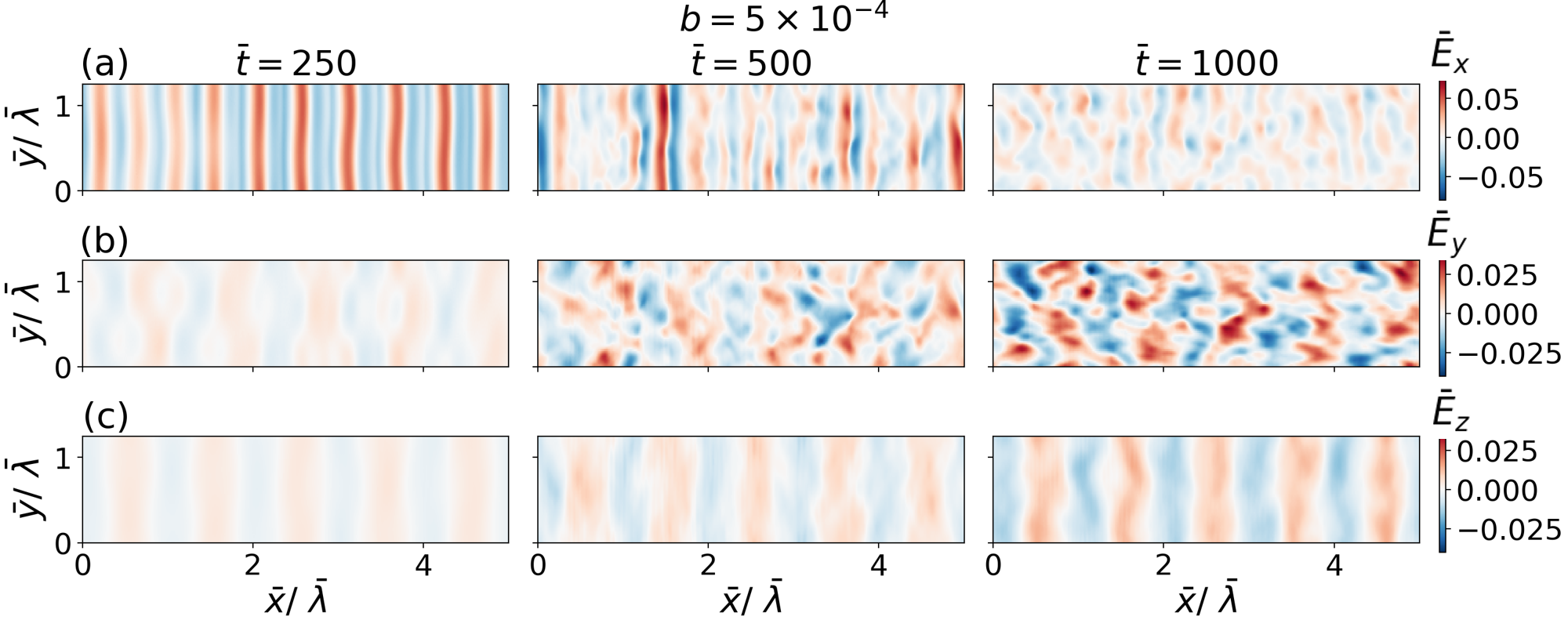}
\caption{\label{fig:E-field_snapshot(b=5e-4)} Snapshots of (a) $\bar{E}_{x}$, (b) $\bar{E}_{y}$, and (c) $\bar{E}_{z}$, for $b=5\times10^{-4}$ case. Colormap indicates the amplitude of electric fields.}
\end{figure}

We next present snapshots of the electric-field profiles for the case $b=5\times10^{-4}$. Figure \ref{fig:E-field_snapshot(b=5e-4)} shows snapshots of (a) $\bar{E}_{x}$, (b) $\bar{E}_{y}$, and (c) $\bar{E}_{z}$, where $\bar{x}=x \ \omega_{\mathrm{c}}/c$ and $\bar{y}=y \ \omega_{\mathrm{c}}/c$. At $\bar{t}=250$, $\bar{E}_{x}$ arises from the relaxation of the bump-on-tail distribution. While some bump electrons are accelerated toward the resonant momentum, others are decelerated and merge with the background population. For $500\lesssim\bar{t}\lesssim1000$, the amplitude of $\bar{E}_{x}$ decreases while its spatial profile remains similar to that at $\bar{t}=250$, continuing to accelerate energetic electrons. During this phase, $\bar{E}_{y}$ and $\bar{E}_{z}$ increase due to the instability itself. 

The most important takeaway is the robustness of the resonant interaction with respect to the spatiotemporal variations of the electric field. The $\bar{E}_x$ structure from the bump-on-tail instability is drastically different from a uniform parallel field assumed in the theoretical analysis. There are also additional perpendicular fields that are generated due to the instability on top of the prescribed R-wave. However, electrons are still prevented from accelerating beyond $\bar{p}_{\textrm{r}}$ as shown in Figure \ref{fig:fe(px,py,t)}(b), indicating that the firewall effect is still robustly in effect. Therefore, our relatively simple theoretical analysis is still applicable to such self-consistent, more complex situations. 

\captionsetup{justification=raggedright,singlelinecheck=false}


\bibliographystyle{jpp}

\bibliography{R-wave}

\end{document}